\documentstyle[11pt]{article} 

\def\be   {\begin{equation}}
\def\ee   {\end{equation}}    
\def\bea  {\begin{array}}     
\def\eea  {\end{array}}       
\def\bef  {\begin{figure}}
\def\ef   {\end{figure}}      
\def\bec  {\begin{center}}
\def\ec   {\end{center}}      
\def\bet  {\begin{tabbing}}
\def\et   {\end{tabbing}}     
\def\beqa {\begin{eqnarray*}}
\def\eeqa {\end{eqnarray*}}   

\newcommand\ds{\displaystyle}
\newcommand\ts{\textstyle}

\newcommand{\ba}{\begin{eqnarray}}
\newcommand{\ea}{\end{eqnarray}}
\newcommand{\om}{\omega} 
\newcommand{\Om}{\Omega}
\newcommand{\re}[1]{\sigma^{(#1)}}
\newcommand{\omr}[1]{\omega^{(#1)}_J}
\newcommand{\mm}[1]{\langle #1 \rangle}
\newcommand{\la}{\langle}
\newcommand{\ra}{\rangle}
\newcommand{\Fi}{\varphi}

      \let\b=\beta               \let\d=\delta  
 \let\f=\varphi           \let\h=\eta      
      \let\l=\lambda              \let\n=\nu      
\let\o=\omega                    \let\s=\sigma   
                    
                  \let\ro=\rho

\let\O=\Omega            
        
\def\Chi{\mathcal{X}}
                \def\cF{{\cal F}}

\newcommand{\RR}{{\mathrm I}\hspace{-0.17 em}{\mathrm R}}
\newcommand{\NN}{{\mathrm I}\hspace{-0.17 em}{\mathrm N}}

\newcommand{\EE}{{\mathrm I} \hspace{-0.21 em} {\mathrm E}}   
\newcommand{\E}{{\mathrm E}}

\renewcommand{\EE}{\E}  

\renewcommand\to{\longrightarrow}
\def\nn {\nonumber}

\newcommand{\de}{\partial }

\def\dif{{\mathrm d }}

\def\le{\left}
\def\ri{\right}
\def\la{\langle}
\def\ra{\rangle}

\def\n #1 {\parallel\! #1 \!\parallel}
\def\ket #1 {|#1\ra}
\def\bra #1 {\la#1|}
\def\mod #1 {|#1|}
\def\pro #1 {\ket{#1} \bra{#1} }
\def\sca #1 #2 {\la{#1} | {#2}\ra }

\newcommand\q[1]{{\mathrm q}_{#1}}

\newcommand\Op[2]{\hat{\mathrm O}^{{}^{(#1)}}_{#2}}

\newcommand{\unsu}[1]{\frac{1}{#1}}

\newcommand{\es}[1]{{\mathrm e}^{#1} }

\newcommand\Fdi[2]{{\mathrm F}_{#1}\le[#2\ri]} 
\newcommand\Ldi[3]{{\mathrm L}_{#1}^{#2}\le[#3\ri]} 
\newcommand\Ff{{\mathrm F}} 
\newcommand\Lf{{\mathrm L}} 

\newcommand\duno[2]{\frac{\d{#1}}{\d{#2}}}

\def\kquater4{k_4}

 \def\spose#1{\hbox to 0pt{#1\hss}}
 \def\ltapprox{\mathrel{\spose{\lower 3pt\hbox{$\mathchar"218$}}
  \raise 2.0pt\hbox{$\mathchar"13C$}}}
 \def\gtapprox{\mathrel{\spose{\lower 3pt\hbox{$\mathchar"218$}}
  \raise 2.0pt\hbox{$\mathchar"13E$}}}
 \def\inapprox{\mathrel{\spose{\lower 3pt\hbox{$\mathchar"218$}}
  \raise 2.0pt\hbox{$\mathchar"232$}}}

\newtheorem{theorem}{Theorem}[section] 
\newtheorem{proposition}[theorem]{Proposition}

\newtheorem{lemma}[theorem]{Lemma}

 \def\today{\number\day \space \ifcase\month\or
 January\or February\or March\or April\or May\or June\or
 July\or August\or September\or October\or November\or December\fi
 \space\number\year}
\newcounter{ore}
\newcounter{minuti}
{\ifnum \value{minuti}>719{\addtocounter{ore}{12}
                           \addtocounter{minuti}{-720}} \fi}
{\ifnum \value{minuti}>359{\addtocounter{ore}{6}
                           \addtocounter{minuti}{-360}} \fi}
{\ifnum \value{minuti}>179{\addtocounter{ore}{3}
                           \addtocounter{minuti}{-180}} \fi}
{\ifnum \value{minuti}>119{\addtocounter{ore}{2}
                           \addtocounter{minuti}{-120}} \fi}
{\ifnum \value{minuti}> 59{\addtocounter{ore}{1}
                           \addtocounter{minuti}{ -60}} \fi}

\begin{document}

\pagestyle{empty}
 \begin{flushright}
 February $14^{\mathrm th}$, 2000  
 \end{flushright}
\vskip 0.5cm
 \centerline{\Large\bf SOME EXACT RESULTS  }
\vskip 0.2cm
 \centerline{\Large\bf ON THE ULTRAMETRIC OVERLAP DISTRIBUTION} 
\vskip 0.2cm
 \centerline{\Large\bf IN MEAN FIELD SPIN GLASS MODELS (I) }
\vskip 1cm
 \centerline{Francesco Baffioni $\mbox{}^1$\hspace{0.5cm}and\hspace{0.5cm}Francesco Rosati $\mbox{}^2$}
\vskip 1cm
\centerline{$\mbox{}^1$ Dipartimento di Matematica}
\centerline{$\mbox{}^2$ Dipartimento di Fisica}
\centerline{Universit\`a di Roma ``Tor Vergata''}
\centerline{Via della Ricerca Scientifica
 1, I-00133 Roma, Italy. }
\begin{center}
\begin{tabular}{rl}
e-mail : & baffioni@mat.uniroma2.it \\[-1.2em]
         & Francesco.Rosati@roma2.infn.it
\end{tabular}\end{center}

 \centerline{} 
\vfill

\centerline{\bf Abstract}
\vskip 0.2cm
The mean field spin glass model is analyzed by a combination of 
mathematically rigororous methods and a powerful {\it Ansatz}. 
The method exploited is general, and can be applied to others 
disordered mean field models such as, {\it e.g.}, neural networks.
 
It is well known that the probability measure of overlaps among replicas
carries the whole physical content of these models. 
A functional order parameter of Parisi type is introduced by 
rigorous methods, according to previous works by F.~Guerra.
By the {\it Ansatz} that the functional order parameter is the correct order 
parameter of the model, we explicitly find the full overlap distribution.
The physical interpretation of the functional order parameter 
is obtained, and ultrametricity of overlaps is derived as a natural
consequence of a branching diffusion process.

It is shown by explicit construction that 
ultrametricity of the 3-replicas overlap distribution 
together with the Ghirlanda--Guerra relations determines the distribution 
of overlaps among $s$ replicas, for any $s$, in terms of 
the one-overlap distribution.

\break

\newpage

\pagestyle{plain}
\setcounter{page}{1}
\baselineskip =16pt


\section{Introduction.}

Mean field spin glass models are considered as a prototype of
disordered, frustrated systems and, more generally, of a large class of 
complex systems that can be successfully analyzed using the ideas developed
in the study of spin glasses~\cite{PMV,Parisi_rev}. Among these, 
the Sherrington--Kirkpatrick model~\cite{SK} has a primary importance.
This model is by now well understood in its general features,
as described by Parisi with an ingenious method and the ultrametric 
{\it Ansatz}~\cite{PMV}. This picture has been confirmed by extensive 
numerical simulations~\cite{num1,Parisi_rev} 
and some rigorous results~\cite{pash,shch,gue1,gue4,tala,AC}. 
In particular, F.~Guerra has given a rigorous 
motivation for the introduction of a functional order parameter of Parisi 
type, and has shown how in this framework a simple {\it Ansatz} allows 
to express the thermodynamic variables and some physical observables in 
terms of that order parameter~\cite{gue1,gue3}. 

In the present paper, the {\it Ansatz} of Guerra is extended, and is 
developed a method to express all physical observables in terms 
of the functional order parameter, in a mathematically rigorous framework.
The method is general, and can be applied to other mean field disordered 
models such as the multi-spin interaction spin glass and the neural networks.

It is well known that the whole physical content of mean field spin 
glass models is contained in the overlap random variables. Given $s$ 
replicas there are $s(s-1)/2$ overlaps between 
them, where $s$ ranges on the natural numbers. Therefore, the physics of the 
model is fully contained in a probability distribution on an 
infinite-dimensional space. 
Overlaps do not fluctuate in the hight temperature phase : the 
Sherrington--Kirkpatrick solution turns out to be 
correct and the overlap 
distribution is trivial. In the low temperature phase this cannot happen : 
overlaps do fluctuate~\cite{PMV,pash,gue1,gue4}.

 Fluctuations are constrained by the 
symmetry under permutations of replicas and by the gauge symmetry. 
Thermodynamical constraints are expressed by Ghirlanda--Guerra relations,
in the slightly stronger case when suitable infinitesimal interactions are
added to the Hamiltonian~\cite{gue4,ghigu} (this is also known as 
the stochastic stability property~\cite{parisi,AC}).
By the {\it Ansatz} that the overlap distribution is ultrametric,
Parisi gave a solution of the model, in terms of a 
functional order parameter~\cite{PMV}.
Ultrametricity is a simple constraint on the support that 
considerably simplifies the overlap distribution : together 
with the previously stated constraints, 
it reduces the problem to the determination of the mono-dimensional,
one-overlap distribution $P_{12}$. This is proven in the last 
section of this paper. 

A functional order parameter of Parisi type can be introduced rigorously
to give a functional representation of the marginal 
martingale function, and therefore of the free energy~\cite{gue1}. 
This representation is not unique : there is an infinite 
set of functional order parameters giving rise to the same free energy.
By an {\it Ansatz} on this representation, some overlap correlation 
functions has been expressed through the functional order 
parameter~\cite{gue3}.
In this paper, by an extension of the {\it Ansatz}, we explicitly find
the full overlap distribution in terms of the functional order parameter, 
and we show how ultrametricity naturally emerges.
  
The method (and the paper) goes as follows. We introduce a 
generating functional of physical observables ({\it i.e.}, 
expectations of overlap 
functions), derived from the marginal martingale function 
(sect.~\ref{sez:generator}). Through the solution of a non-linear 
antiparabolic equation, and exploiting the {\it Ansatz}, we represent it in 
terms of the functional order parameter $x$ (sect.~\ref{sez:rapp}). Then, we 
solve the antiparabolic equation by asymptotic expansion and explicitly 
find the overlap probability distribution. 
The physical interpretation of the functional order parameter 
is obtained, and ultrametricity of overlaps is derived as a natural
consequence of the branching diffusion process underlying the equation 
(sect.~\ref{sez:antip}).

Finally, it is shown that complete ultrametricity of overlaps results
from ultrametricity of the  3-replicas overlap distribution. 
Moreover, it is proved that ultrametricity and the Ghirlanda--Guerra 
identities are
{\it complementary} in order to determine the full overlap distribution, 
in the sense that one can hold independently of the other, but together 
they determine explicitly the overlap measure in terms of the one-overlap 
distribution~$P_{12}$ (sect.~\ref{sez:ultima}).


\section{Overlaps in the Sherrington--Kirkpatrick model.} 
\label{sez:over}

The mean field model of a spin glass is defined on sites $i=1,2,\ldots ,N$. To
each site is assigned the Ising spin variable $\s_i = \pm 1$, so that a
configuration of the system is described by the application 
$\s :\; i \rightarrow \s_i \in Z_2 = \{ -1,1\}$. The spins on two different
sites $i$ and  $j$ are coupled through the random variables $J_{ij}$, all
independent from each other and equally distributed. For the sake of 
simplicity we assume a Gaussian distribution, with
\be
\E(J_{ij}) = 0, \;\;\; \E(J_{ij}^2) = 1,
\ee
where $\E$ denotes averages on the $J$ variables. The $J_{ij}$'s are called
{\it quenched} variables, because they do not participate to thermalisation. 
The Hamiltonian of the Sherrington--Kirkpatrick model is
\be
H_N(\s,J) = - \frac{1}{\sqrt{N}} \sum_{(i,j)} J_{ij} \s_i \s_j ,
\ee
where the sum extends over all the $N(N-1)/2$ couples of sites. The
normalization factor $1/\sqrt{N}$ is needed to have the correct behavior 
of the thermodynamic variables in the limit $N\rightarrow \infty$. Denoting 
with $\beta$ the inverse temperature (in proper units), we introduce the 
partition function $Z_N (\beta,J)$ and the free energy 
density $f_N (\beta,J)$ :
\ba
Z_N (\beta,J) &=& \sum_{\s_1 \ldots \s_N} e^{-\beta H_N (\s,J)},\\
-\beta f_N (\beta,J)  &=& \frac{1}{N} \log Z_N (\beta,J) .
\label{enlib}
\ea
The associated Boltzmann state $\om_{N,\beta,J}$ is defined by
\be
\om_{N,\beta,J}(A) = \frac{1}{Z_N (\beta,J)} \sum_{\s_1 \ldots \s_N}
                     A(\s) e^{-\beta H_N (\s,J)} ,
\label{boltz}
\ee
for a generic function $A$ of the spin variables. Another relevant quantity is 
the average of internal energy density $u_N (\beta)$
\be
u_N (\beta) = \frac{1}{N} \E\,\om_{N,\beta,J} (H_N (\s,J)) =
\E \frac{\de}{\de\beta} (\beta f_N (\beta,J) )
.
\label{enint}
\ee

In the thermodynamic limit the free energy density is self-averaging in 
quadratic mean~\cite{pash}. For the internal energy density the same 
property has been proven for almost all values of $\beta$, but is believed 
to hold without restrictions~\cite{gue4} . 
 
One of the main features of the mean field spin glass model is the existence
of observables that do not self-average in the thermodynamic limit. This
is one of the fundamental intuitions contained in the Parisi {\it Ansatz}
of replica symmetry breaking. 
Indeed Pastur and Shcherbina have proven that if
a suitably chosen order parameter
(coming from the response of the system to an external random field)
is self-averaging in the thermodynamic limit,
then the solution of the model has the Sherrington--Kirkpatrick 
form~\cite{pash,shch}:  
this is unphysical at high $\beta$, because it gives negative entropy. 
Moreover, self-averaging of the Edward--Anderson order
parameter implies that the overlap 
distribution is the trivial one corresponding to the replica 
symmetric {\it Ansatz} of S.--K.~\cite{gue4}.

Let us consider $s$ copies (replicas) of the system, whose configurations
are given by the Ising spin variables $\re{1}_i, \ldots ,\re{s}_i$, and
denote with $\omr{a}, a=1,2,\ldots,s $ the relative Boltzmann states,
the dependence on $\beta$ and $N$ being understood. We introduce the product
state $\Om_J$ by
\be
\Om_J = \omr{1} \omr{2} \cdots \omr{s}
\label{pinna}
,\ee
where all the states $\omr{a}$ are subject to the same values of the
quenched variables $J$, and the same temperature $\beta$.

The overlap between the two replicas $a$ and $b$,  $Q_{ab}$,
is defined by
\be
Q_{ab} = \frac{1}{N} \sum_i \re{a}_i \re{b}_i
,\ee
with the obvious bounds $-1 \leq Q_{ab} \leq 1$. 

The importance of overlaps lies in the fact that all physical observables 
 can be expressed in the form 
\be
\E\Om_J [ F(Q_{12}, Q_{13}, \ldots)],
\ee
for some function $F$. For $F$ smooth, we can introduce the 
random variables $q_{12},q_{13},\ldots$,
through the definition of their averages
\be
\mm{F(q_{12},q_{13},\ldots)} = \E \Om_J [ F(Q_{12},Q_{13},\ldots) ]
\label{pizza}
.\ee
Notice that the expectation $\mm{\cdot}$ includes
both the thermal average and the average $\E$ over disorder.
The overlap distribution carries the whole physical content of the 
model~\cite{gue4}.

Let us recall some considerations about the overlap distribution.
The average $\E$ over quenched variables introduces correlation between
different groups of replicas. For example we have, in general,
\be
\mm{q_{12}^2 q_{34}^2} \neq \mm{q_{12}^2} \mm{q_{34}^2}
.\ee

The $\mm{\cdot}$ average is obviously invariant under permutations of  
replica
indices (e.g.~$\mm{q_{12}^2 q_{13}^2} = \mm{q_{23}^2 q_{13}^2}$,
$\mm{q_{12}^2} = \mm{q_{34}^2}$). Moreover, it is invariant under
the gauge transformations defined by
\be
q_{ab} \longrightarrow \varepsilon_a  q_{ab} \varepsilon_b 
,\ee
where $\varepsilon_a = \pm 1$. 
This is an easy consequence of the fact that each
of the $\omr{a}$ is an even state on the respective $\re{a}$. It follows, for
instance, that polynomials in the overlaps which are not gauge invariant have
zero mean. These symmetries furnish important restrictions on the the overlap
distribution, but even more important constraints have been given 
by~\cite{gue4,ghigu}, using simple arguments based on convexity properties and
positivity of fluctuations.
Consider $s$ replicas, and the $s(s-1)/2$ overlaps between them. Let us 
denote by ${\mathcal{A}}_s$ the associated algebra of observables. 
Introduce the overlap $q_{a,s+1}$, between replica $a$ and an
additional replica $s+1$, and consider the conditional probability
distribution $\tilde{P}_{(a,s+1)}(q_{a,s+1}|{\mathcal{A}}_s)$ of 
$q_{a,s+1}$ given the overlaps among the first $s$ replicas.
By adding to the Hamiltonian suitable infinitesimal
external fields, and taking the thermodynamic limit with a careful
procedure, Guerra and Ghirlanda have demonstrated that the following theorem 
holds for a very general class of probability measures, including short 
range models~\cite{ghigu}.  
\begin{theorem}
\label{thegg} 
Given the overlaps among $s$  replicas, the overlap between one of these,
let say $a$, and an additional replica $s+1$ is either independent of the 
former overlaps, or it is identical to one of the overlaps $q_{ab}$, with $b$ 
running from 1 to $s$, excluding $a$. Each of these cases 
have probability $1/s$:
\ba
\tilde{P}_{(a,s+1)} ( q_{a,s+1} |{\mathcal{A}}_s )
= \frac{1}{s} P_{12} (q_{a,s+1}) + \frac{1}{s} \sum_{b \not = a}
\delta (q_{a,s+1} - q_{ab} ) 
.\label{prima}
\ea

\end{theorem}
 
Results of this kind have been obtained by Parisi in the frame of replica 
method~\cite{parisi}, and by Aizenmann and Contucci~\cite{AC}.

\section{A generator of overlap distributions.}
\label{sez:generator}

Let $\o$ be a generic even state on the Ising spins $\s_1,\ldots,\s_N$, 
possibly depending on the quenched variables $J_{ij}$ and let 
$ f_{{}_1}\ :\,\RR \longrightarrow \RR $ be an 
even, convex  function, such that $|f_{{}_1}(y)| \leq c|y|$ asymptotically with
$|y|\rightarrow\infty$ for some positive $c$. Let the generating functional 
$\psi_N (\o,\,f_{{}_1})$ be given by 
\be
\psi_N (\o,\,f_{{}_1}) = \EE \log\o ( \exp f_{{}_1}(h_N(\s,J)))
,\ee
where  $h_N(\s,J)= N^{-1/2} \sum_i J_i \s_i$ is the cavity field, and the 
$J_i$'s are fresh noise with the same properties of $J_{ij}$.
The functional $\psi_N (\o,\,f_1)$ contains all informations on the 
distribution of the replicated cavity fields 
$h^{(a)}\equiv h_N(\re{a},J)$. That, in turn, 
is related to the overlap distribution through the well known formula
\be
\EE\Omega_J \left( \exp\left({\ts i\sum_a} k_a h^{(a)} \right) \right)
= \left\la \exp\left( {\ts -\sum_{a,b}} k_a k_b q_{ab}/2 \right) \right\ra
\label{proprio}
.\ee
We expand the logarithm in power series and we introduce replicas:
\be 
 \bea{lcl} 
 \psi_N (\o,\,f_{{}_1}) &=& {\ds \EE \le[\ln \le(1-\o
                        \le(1-\exp\le(f_{{}_1}( h )\ri)\ri)\ri)  
                                                  \ri] } \\ 
            &=& {\ds - \sum_{s=1}^\infty \frac{1}{s} 
                \EE \left[\o \left(1-\exp \le(f_{{}_1} ( h )\ri) 
                                       \right)\right]^s}  \\
            &=& {\ds - \sum_{s=1}^\infty \frac{1}{s} 
                \EE\O_J \left[\prod_{a=1}^s
                        \le(1-\es{f_{{}_1}(h^{(a)})}\ri)\ri]} 
 \eea 
\label{eq:1} 
\ee 
where  $h$ denotes the cavity field and $h^{(a)}$ its replicas. 
\par\noindent
Let us introduce the generalized Fourier transform $\phi$, 
which is a well defined even generalized function~:
\be
  1-\exp \le( f_{{}_1}(y) \ri) = \int_{-\infty}^\infty
	 \dif k \, {\phi}(k) e^{iky }
\ee                                    
\par\noindent
By the convenient replacement $\varphi(k) \equiv  \phi (k)\exp (-k^2/2)$, 
we finally have
\be 
  \psi_N(\o,\,f_{{}_1}) = {\ds - \sum_{s=1}^\infty \frac{1}{s} 
               \int \dif^s k \prod_{a=1}^s {\f}(k_a) 
                \le\la \exp \le( 
               {\ts - \sum_{(a,b)} k_a k_b \q{ab} }\ri)\ri\ra} 
\label{eqmart1} 
\ee 
where the sum in the exponential is over the couples $(a,b)$, for 
$1\leq a < b \leq s$. The dependence of the r.h.s. on $f_1$ is through the 
function $\varphi$. Notice that  terms $s=2,3$ contain the characteristic 
functions of the distributions of overlaps among 2 and 3 replicas, 
respectively. 

It important to notice that the 
thermodynamic functions can be represented through the 
functional~$\psi_N(\o,\,f_1)$.
Consider the case $f_1(y) = \log\cosh \beta y$, and the corresponding
function $\psi_N^\star(\beta) \equiv \psi_N(\o,\,\log\cosh\beta\,\cdot)$,
where $\omega$ is the Boltzman state of SK model. Then the following 
holds~\cite{gue1}.
\begin{proposition}
Assume the existance of the limit $\lim_{N\rightarrow\infty} 
\psi_N^\star(\beta) \,=\, \psi^\star(\beta)$, uniformly on a compact region
$0 \leq \beta\leq\tilde{\beta}$, with $\psi^\star$ continuous in $\beta$, 
as a consequence. Let us define
\be
\alpha(\beta) = \log 2 + \int_0^1 \psi^\star (\beta \sqrt{1-q})\ dq
,\ee
so that the $\beta$ derivative $\alpha'(\beta)$ exists and the following 
holds
\be
\alpha(\beta) + \beta\alpha'(\beta)/2 = \log 2 + \psi^\star(\beta)
.\ee
Then, we have, for $0 \leq \beta\leq\tilde{\beta}$,
\be
\lim_{N\rightarrow\infty} \frac{1}{N} \E \left(\log Z_N(\b,\,J)\right)
 = \alpha(\b)
,\ \
\lim_{N\rightarrow\infty} \frac{1}{N} 
\partial_\b \E \left(\log Z_N(\b,\,J)\right) = \alpha'(\b)
.\ee
\end{proposition}


\section{The functional order parameter.}
\label{sez:rapp}

In the frame of the cavity method, a functional order parameter of Parisi 
type was introduced by Guerra as a functional representation of the marginal 
martingale 
function~\cite{gue1}. Then, he showed that  by a simple {\it Ansatz}
some 
overlap correlation functions can be expressed in terms of the functional 
order parameter~\cite{gue3}.

In this section we give an extension of the representation Theorem, thus 
obtaining a functional representation of the physical observables. Exploiting 
the {\it Ansatz}, the generating functional $\psi_N(\om,\, f_1)$ is 
expressed in terms of the functional order parameter. Therefore, the 
explicit form of the overlap distribution can be extracted. 

Let us introduce the convex set $\Chi$ of functional order parameters of the 
type
\be
x : \; [0,1] \ni q \longrightarrow x(q) \in [0,1]
,\ee
with the $L^1(dq)$ distance norm. We induce on $\Chi$ a partial ordering, by
defining $x\leq \bar{x}$ if $x(q)\leq \bar{x}(q)$ for all $0\leq q \leq 1$, and
introduce the extremal order parameters $x_0(q) \equiv 0$ and $x_1(q)\equiv 1$,
such that for any $x$ we have $x_0(q) \leq x(q) \leq x_1(q)$.

For each $x$ in $\Chi$, and for suitable $f_{{}_1}$ 
(see the previuos section), 
let us define the function with values $f(q,y;\,x,f_{{}_1}),
0\leq q \leq 1 ,\;  y\in \RR$, as the solution of the nonlinear
antiparabolic equation
\be
\de_q f + \frac{1}{2} (f'' + x(q)f'^2) = 0
,\label{antip}
\ee
with final condition
\be
f(1,y;\,x,f_{{}_1}) = f_{{}_1}(y)
\label{finale}
\ee
In (\ref{antip}), $f'=\de_y f$ and $f''=\de_y^2 f$.

With these definitions, the following representation theorem holds~\cite{gue1}.
\begin{theorem}
\label{th:rapp}
There exists a nonempty hyper-surface $\Sigma_N(\om,f_{{}_1})$ in $\Chi$ such
that, for any $x\in \Sigma_N(\om,f_{{}_1})$ and $f$ solution
of~(\ref{antip},\ref{finale}), we have the following representation
\be
\psi_N (\om,\, f_1) = f(0,0;\,x,f_1)
.\label{rappr}
\ee
\mbox{}\end{theorem}

Any family of functional order parameters, $x_\epsilon$, depending continuously
in the $L^{{}^1}$ norm on the variable~$\epsilon$, $0\leq\epsilon\leq 1$, with
$x_0\equiv 0$, and $x_{{}_1}\equiv 1$, and nondecreasing in $\epsilon$, must
necessarily cross $\Sigma_N(\om,f_{{}_1})$ for some value of the variable
$\epsilon$ (we say that $\Sigma_N(\om,f_{{}_1})$ has the monotone intersection
property). A similar representation holds also in the infinite volume limit.

Of particular interest are those states $\om$ such that the
representation~(\ref{rappr}) holds with some $x$, depending on $\om$, but
independent on $f_1$, with some possible error vanishing in the
limit~$N\rightarrow\infty$. We call such states $x$-representable. Some
examples of $x$-representable states are shown in \cite{gue3}.

An attractive conjecture is that the Boltzmann state of  mean 
field spin glass models is $x$-representable. Indeed, this must be the case 
if $x$ is the correct order parameter. We will refer to this as the 
{\it tomographic Ansatz} : in the $\Chi$ space the hyper-surfaces 
 $\{ \Sigma_\infty (\o, f_1), f_1 \in F_1 \}$ have a common point $x$, 
which gives the physical content of the theory.
By this {\it Ansatz}, we can express the full probability distribution
of  overlaps in terms of the functional order parameter.
Let us state the following theorem, one of the main results of this paper, 
leaving the proof to the next sections.

\begin{theorem}
\label{th:dist}
Let $\om$ be an even state on the Ising spin variables $\s_i$,
depending on the quenched variables $J$, and suppose it is $x$-representable,
with $x(0)=0$ and $x(1)=1$. Then the following holds. 
\begin{enumerate}
\item [a)]
The probability distributions
of overlaps among $s=2,3  $ replicas are given  in terms of
the functional order parameter $x$ by the
 following expressions :
\ba
P_{12}(q)\equiv P(q) &=& \frac{d}{dq} x(q)  ,\label{pidue}\\
P_{12,23,13} (q_{12},\,q_{23},\,q_{13}) &=& \frac{1}{2} x(q_{12}) P(q_{12})
\delta(q_{12} - q_{23}) \delta(q_{12} - q_{13}) +   
\nonumber\\
&&
\!\!\mbox{} + \frac{1}{2}\left( P(q_{12}) P(q_{23}) \theta(q_{12} - q_{23})
\delta(q_{13} - q_{23}) + \mathrm{cyclic\ perm.} \right) 
.\label{pitre}
\ea
\item [b)]
Assume in addition the hypothesis of Theorem~\ref{thegg}.
Then the overlap distribution is uniquely determined in terms of the 
functional order parameter $x$, and the $s$-replicas marginals 
({\it i.e.}, the distribution of overlaps among $s$ replicas) 
can be given explicitly for any~$s$ (see section~\ref{sez:ultima}).
\end{enumerate}
\end{theorem}
We have used Dirac's $\delta$ function and the step function $\theta$.
Extension to regions of negative $q$'s is made by gauge symmetry, as shown 
in the next section.
Equation (\ref{pidue}) gives the physical meaning of the functional order 
parameter; equation (\ref{pitre}) corresponds to ultrametricity of the overlap 
distribution, as is proven in the following.
For other values of $x(0)$ and $x(1)$,  slightly different results can be 
obtained. 

As is shown extensively in the next section, ultrametricity arises naturally
as a consequence of the branching diffusion process underlying 
equation~(\ref{antip}, \ref{finale}).

All results are in full agreement with those found in the frame of replica
symmetry breaking method with Parisi {\it Ansatz}~\cite{PMV}.


\section{Asymptotic solution of the antiparabolic equation.} 
 \label{sez:antip}

The results (\ref{pidue},\ref{pitre}) of Theorem~\ref{th:dist}
are obtained by equation (~\ref{rappr}), and the tomographic {\it Ansatz}.
Both members of equation (\ref{rappr}) are expressed as asymptotic series,
which are then compared term by term. The first one is given by 
equation (\ref{eqmart1}), the second is obtained in this section.
  
Let us transform  equation (\ref{antip}, \ref{finale}) 
$$ 
 \le\{ 
  \bea{l} 
   {\ds \de_q\ f_q + \frac{1}{2} \le( f_q'' + x_{q}\ {f_q'}^2\ri) = 0} \\ 
   {\ds f(1,y;x,f_{{}_1}) = f_{{}_1}(y)} 
  \eea 
 \ri.
$$  
into an equivalent form. When $x(0) = 0$ and $x(1) = 1$, satisfied by physical 
order parameters, it is convenient to make the substitution
\be 
 g_q\,(y) = \le[ 1 - \exp \le( x_q\ f_q(y)  \ri) \ri] /x_q 
\ee 
the $x$ and $f_{{}_1}$ dependence of $f$ being understood. The resulting 
equation is 
\be 
 \le\{ 
  \bea{l} 
   {\ds \de_q\ g_q + \frac{1}{2}\ g_q''  = 
    \ro_q \le[ x_{q}\ g_q + (1-x_{q}\ g_{q}) \ln(1-x_{q}\ g_{q}) \ri]
                                                           /\ x_{q}^2} \\ 
   {\ds g(1,y;x,f_{{}_1}) \equiv g_{{}_1}(y) = 1-\exp \le( f_{{}_1}(y)\ri)} 
  \eea 
 \ri.
\label{hjk}
\ee 
where $x_q\equiv x(q)$ and 
$\ro_q\equiv \dif x_q / \dif q $. Notice that the final condition 
for  $g$ is equal to the function used in the expansion 
of $\psi_{N}(\o,f_1)$ (eq.~\ref{eq:1}), and that $g(0,y)\ =\ f(0,y) $. 
Let us re-write equation (\ref{hjk}) in integral form:  
\be 
   g_q = N_{1-q}\ast 
                 g_1 + \int_q^1 \dif q'\  
          \frac{\ro_{q'}}{x_{q'}^2}\ N_{q'-q} \ast 
          \le[x_{q'}\ g_{q'}+(1-x_{q'}\ g_{q'})\ln(1-x_{q'}\ g_{q'}) \ri] 
\label{sonio} 
\ee 
as one can straightforwardly see by simple inspection. Here 
$ N_q\equiv N(q,y)=\exp\le( -y^2/2q\ri)/\sqrt{2\pi q}$ 
is the usual heat kernel and the symbol $\ast$ is the convolution operation 
on $y$ variable
\footnote{$(f\ast g)(y)\equiv 
          \int\dif y' f(y-y')\ g(y')$}. 

Equation (\ref{sonio}) can be handled by asymptotic expansion of the
r.h.s. term under square brackets~:
\be 
  g_q =   N_{1-q}\ast g_1 + \sum_{i=2}^\infty \frac{1}{i(i-1)} 
          \int_q^1 \dif q'\  
          \ro_{q'}\ x_{q'}^{(i-2)}\ N_{q'-q} \ast \le[\le(g_{q'}\ri)^i \ri] 
\label{eq:14}
\ee 
We write the above equation in the ``moment space''~: let $\h_q$ be 
the Fourier transform of $ N_q\ast g_q$ in the $y$ variable 
and $\f$ that of $N_1\ast\ g_{{}_1}$.
Thus we have, after simple algebraic manipulation 
\be 
 \h_z = \f_z + \Fdi{z}{\h}  
 \label{uno} 
\label{fund}
\ee 
where $z$ is a collective variable for $(q,k)$, 
$\f_z\equiv\f(k)$ is the same function appearing in eq.~(\ref{eqmart1}),
$\Fdi{z}{\h}$ is a function of $z$ and a functional of $\h$
\be 
 \Fdi{z}{\h}\equiv 
 \sum_{i=2}^\infty\unsu{i!}\ \Op{i}{z}\le[\h,\dots,\h\ri] 
\ee 
and the $\Op{i}{z}$ are well defined multi-linear integral operators
$$ 
 \bea{rl} 
  {\ds\Op{i}{z} \le[{\f}_{{}_1},\dots, {\f}_i\ri]} & \\ 
  {\ds\equiv (i-2)!} &{\ds \int \dif^i k\ \d(k_1+\dots+k_i-k)\  
              \prod_{a=1}^i\ {\f}_a(k_{a})} \\
             &{\ds \int_q^1 \dif q' \ro_{q'}\ x_{q'}{}^{i-2}\  
              \exp(-q'\sum_{(a,b)} k_a k_b)} 
 \eea 
$$ 
Every term in the asymptotic expansion is well defined.   
Notice that the representation Theorem~\ref{th:rapp} can be 
rephrased as 
\be
     \psi_N(\o,{f_1}) = - g(0,0) = - \int \dif k\ \h (0,k)
\label{eq:fin}
\ee
and this is the form that we will use in the sequel. 

The $i$--th functional derivative of F${}_z [ \h ]$  
w.r.t. $\h$ calculated in zero gives the integral kernel of the 
$\Op{i}{z}$ operator. In particular 
\be 
 \le.\Fdi{z}{\h}\ri|_{\h=0} = 0 \ \ \ \ ; \ \ \ \
 \le.\frac{\d\Ff_z\le[\h\ri]}{\d\h_w}\ri|_{\h=0} = 0 
 \label{su}
\ee
Replacing ${\h}=\Lf[\f]$ in (\ref{uno}) we have 
\be 
 \Ldi{z}{}{\f} \equiv \f_z + \Fdi{z}{\Lf[\f]},\ \ \ \ \ \
 \label{ide}
\ee 
which defines iteratively 
the inverse functional $L_z\ [\ . \ ]$~: 
\be 
 \Ldi{z}{}{\f} =
  \sum_{s=1}^\infty \unsu{s!} 
   \int \dif^i k\ \prod_{a=1}^s{\f}(k_{a})\ \ L_z^{(s)} (k_1,\dots,k_s)
\ee 
It is easy to check that 
\be 
 \le.\Ldi{z}{}{\f}\ri|_{\f=0} = 0  \ \ \ \ ; \ \ \ \
 \le.\duno{\Ldi{z}{}{\f} }{\f_w}\ri|_{\f=0} = \d_{z-w} 
 \label{giu} 
\ee 
where $\d_{z-w} $ is the usual Dirac's function,
and that, by derivating (\ref{ide}) w.r.t. $\f$,   
\be 
 \duno{\Lf_z\le[\f\ri]}{\f_w} \equiv \d_{z-w} + 
       \int\dif w' \duno{\Ff_z}{\h_{w'}}\le[\Lf[\f]\ri]\ 
                   \duno{\Lf_{w'}\le[\f\ri]}{\f_w} 
 \label{allgraf} 
\ee 
Subsequent functional derivatives w.r.t. $\f$, calculated in $\f\equiv 0$, 
and the properties (\ref{su}) and (\ref{giu}) allow us to obtain 
straightforwardly all the integral kernels 
$L_{z}^{(s)}(k_1,\dots,k_s)$ for any $s$, in terms  
of $\Op{i}{z}$ operators. We thus obtain 
\be 
 \int\dif k\, \h (0,k) = \sum_{s=1}^\infty 
     \frac{1}{s}\int\dif^s k \prod_{a=1}^s\f(k_a) 
     \frac{1}{(s-1)!}{\Lf}_{o}^{(s)}(k_1,\dots,k_s)  
 \label{etasvil}
\ee 
where ${\mathrm L}_{o}^{(s)}(k_1,\dots,k_s)$ are 
the integral kernels, calculated for $q=0$ and     
integrated on the overall delta dependence in 
the $k$ variable. They are of the form:
\be 
 \frac{1}{(s-1)!}{\Lf}_{o}^{(s)}(k_1,\dots,k_s) = 
 \int\prod_{(a,b)}\dif {\mathrm y}_{ab}\  
     \rho_s^{ {}^{(+)}}(\{ {\mathrm y}_{ab}\}) \exp\le({\ts - \sum_{(a,b)} 
     k_ak_{b}{\mathrm y}_{ab}}\ri)
 \label{ldef} 
\ee
where $\rho_s^{{}^{(+)}}$ has support on a subset of $[0,1]^{s(s-1)/2}$.
In appendix we report the explicit expressions of $\rho_s^{{}^{(+)}}$, 
for $s=2,3,4$ and the recipe to construct it for a generic $s$.  

By eq.~(\ref{eq:fin}) we can compare the asymptotic expansions in 
eq.~(\ref{eqmart1}) and eq.~(\ref{etasvil}).
As the function $\f(k)$ is even, only the even part in the $k$'s of 
integral kernels from both sides can be equated. 
Let us define on $[-1,1]^{s(s-1)/2}$ the function $\rho_s$, 
extending by 'gauge symmetry' the  function $\rho_s^{{}^{(+)}}$ 
\begin{eqnarray}
  \rho_s(\{{\mathrm y}_{ab}\}) 
  &=& 2^{-s} \sum_{\{\varepsilon\}} \rho_s^{{}^{(+)}}
      (\{\varepsilon_a{\mathrm y}_{ab}\varepsilon_b\}) \nn\\
  &=& 2^{-(s-1)} \sum_{\{\varepsilon\}:\varepsilon_1=1} 
      \rho_s^{{}^{(+)}} (\{\varepsilon_a{\mathrm y}_{ab}\varepsilon_b\}) 
\end{eqnarray}
where the sums run over all the $\varepsilon_a = \pm 1 ,\ a=2,\ldots,s$ and 
$\rho_s^{{}^{(+)}} = 0$ if its argument is outside 
$[0,1]^{s(s-1)/2}$; we have used the invariance 
$\varepsilon_a\to -\varepsilon_a$ to fix $\varepsilon_1=1$ in the second line.
We finally have 
\be
   \le\la \exp\le({\ts -\sum_{(a,b)} 
       k_ak_b{\mathrm q}_{ab}}\ri)\ri\ra = 
   \int\prod_{(a,b)}\dif {\mathrm y}_{ab}\  
       \rho_s(\{ {\mathrm y}_{ab}\}) \exp\le({\ts -\sum_{(a,b)} 
       k_ak_{b}{\mathrm y}_{ab}}\ri)  
 \label{fine1}
\ee
which proves part $a$ of Theorem 4.2. 

For $s\ >\ 3$ the l.h.s. of eq.~(\ref{fine1}) does not correspond to the 
characteristic function of the overlap distribution, as the number of the
$k$'s parameters is not sufficient, but to its restriction on a 
hyper-surface of dimension $s$.
In the next section, assuming the hypothesis of 
Theorem~\ref{thegg},
we show that the results obtained so far allow us to 
construct the full overlap distribution function. 
The resulting $s=4$ overlap distribution coincides with $\rho_s$. 
This is a strong indication that $\rho_s$ is the correct distribution also in 
the case of no additional interactions, as required by Theorem~\ref{thegg}. 

For a generic $s$ the distribution $\rho_s$
has the ultrametric form  
\be
 \rho_s(\{ {\mathrm y}_{ab} \} ) = 
 \sum_{i:A_i^s\subset A^s} p_i\   
       \rho_s^{{}^{(i)}}(\{ {\mathrm y}_{ab}\}|A_i^s)\   
 \label{fine2}
\ee
Here $A_i^s$ are disjoint sets, made by portions of 
hyper-planes in $[-1,1]^{s(s-1)/2}$, with dimension $|A_i^s|\leq s-1$;
$A^s$ is the union set; $p_i$ are positive numbers,
which sum to one, and  $ \rho_s^{{}^{(i)}}(\cdots|A_i^s)$ 
are probability densities, whose supports are the sets 
$A_i^s$. The r.h.s. can thus be interpreted as a composite probability
formula: $p_i$ is the probability that the {\it ultrametric event} $A_i^s$
happens and $\rho_s^{{}^{(i)}}(\cdots |A_i^s)$ is the {\it overlap} 
probability, conditioned to $A_i^s$. 
The events $A_i^s$ are disjoint and each $\rho_s^{{}^{(i)}}$
effectively depends only on at most $s-1$ variables.


\section{Ultrametric distributions.}
\label{sez:ultima}

Consider the set $\Phi$ of random variables $q_{a,b} \in [-1,\,1]$ :
\be
\Phi = \{ q_{a,b} \, , \;\, (a,b)\in\Fi \subset C \}
,\ee
where $C$ is the set of couples $(a,b)$ of natural numbers $a,b\in \NN ,\
a < b$. To the set $\Fi$ are then associated a probability space and the 
probability distribution
$P_\Fi (\Phi)$ on it. The distribution functions $P_\Fi$ satisfy the
consistency conditions
\be \int P_{\Fi,\Fi'} (\Phi,\,\Phi') \, \prod_{\alpha\in\Fi'} dq_\alpha
= P_\Fi (\Phi)
,\label{coarse}
\ee
for all disjoint sets $\Fi,\Fi'\subset C$. In the following we will often write
$\{A,B\} \equiv A\cup B$ and $ab \equiv (a,b)$
when not ambiguous. Let us introduce the operator $\Pi_{l,m}$ that, acting on
$\Fi$, permutates the indices $l$ and $m$. E.g.:
$\Pi_{12} \{ (1,2),\, (2,3)\} = \{(1,2),\,(1,3) \}.$
Let $\Fi' = \Pi_{l,m} \Fi$, and denote  by 
$\Phi'$  the associated set of $q$'s. According to the symmetries of the
$\mm{ - }$ average, we ask the probability measure $P$ to be 
gauge invariant, and symmetric under  permutations of indices, 
in the following sense:
\be
P_{\Fi'} (\Phi') = P_\Fi (\Phi) = P_\Fi ( \Pi_{l,m} \Phi' )
\equiv \left( \Pi_{l,m} \, P_\Fi \right) \, (\Phi')
.\ee
This defines the operator $\Pi_{l,m}$ on the space of distributions.

Let us consider a very particular class of distributions for the overlaps
between three replicas, {\it i.e.}, the ultrametric distributions :
\ba
P_{12,23,13} (q_{12},\,q_{23},\,q_{13}) &=&
       	B(q_{12}, q_{23})\, \theta (q_{12} - q_{23})\, \delta (q_{13} - q_{23})
\nonumber\\
&+&	B(q_{23}, q_{12})\, \theta (q_{23} - q_{12})\, \delta (q_{13} - q_{12})
\label{treultra}\\
&+&	B(q_{13}, q_{23})\, \theta (q_{13} - q_{23})\, \delta (q_{12} - q_{23})
,\nonumber
\ea
where $B$ is a distribution.
This simply states that among the three overlaps, two are
equal and the third is greater or equal. From eq.~(\ref{treultra}), a
simple application of symmetries and of the consistency conditions
 (\ref{coarse}), leads to

\begin{proposition} \label{teo1} 
If the distribution $P_{12,23,13}$ has the form (\ref{treultra}), then for
any tern of replicas, $(a,b,c)$, the operator $\cF_{a,b,c}$ is defined 
such that
\be
P_{ab,ac,\Fi} = \cF_{a,b,c} ( P_{ab,\Fi} ,\, P_{ac, \Fi} )
,\label{doppia}
\ee
where $\Fi \subset C ,\ (b,c)\in\Fi$ and $(a,b),\,(a,c) \not\in\Fi$. 
The operator $\cF_{a,b,c}$ is defined through its values
\ba
&&\cF_{a,b,c} ( P_{ab,\Fi} ,\, P_{ac, \Fi} ) 
\, (q_{ab},\,q_{ac};\,\Phi) = \nn\\
&=& P_{ab,\Fi} (q_{ab};\Phi) \, 
	\left[ \theta (q_{ab} - q_{bc}) \delta (q_{ac} - q_{bc}) +
	\theta (q_{bc} - q_{ab}) \delta (q_{ac} - q_{ab})  \right] +
\nonumber\\
&+&
P_{ac,\Fi} (q_{ac};\,\Phi)\, \theta (q_{ac} - q_{bc}) \delta (q_{ab} - q_{bc})
\nn\\
&-&
\delta(q_{ab} - q_{bc}) \delta(q_{ab} - q_{ac})
\int P_{ac,\Fi} (q_{ac};\,\Phi) \theta (q_{ac} - q_{bc}) \,dq_{ac}
\label{ciuppi}.
\ea
\end{proposition}

Note that when the set $\Fi$ is symmetric under  permutation of 
indices $b,c$, we can introduce the operator $\widetilde{\cF}_{a,b,c}$, 
\be
P_{ac,\Fi'} = \widetilde{\cF}_{a,b,c} (P_{\Fi'}) \equiv
\cF_{a,b,c} (P_{\Fi'},\, \Pi_{b,c} P_{\Fi'} )
,\label{singola}
\ee
where $\Fi'\equiv\{(a,b),\,\Fi\}$.

This property of the overlap distribution corresponds to ultrametricity.
In fact, eq.~(\ref{ciuppi}) simply states that for {\it any} triangle of 
overlaps in a given set $\tilde\Fi$, two overlaps are equal and the third
is greater or equal. 
The proof of the theorem and of the subsequent lemma as well, does not depend 
on the nature of the $q_{\cdot,\cdot}$ variables, but only on symmetries and 
general properties of probability spaces.

\begin{lemma}\label{lemma}
In the hypothesis of theorem \ref{teo1}\ we can express the probability
distribution of the overlaps between $s+1$ replicas in terms of the
distribution of the overlaps between $s$ replicas and $q_{1,s+1}$ (for
$s\geq3$).
\end{lemma}

The proof goes as follows. Given $s\geq3$ and $l\leq s$,
we define the set $\Fi_{s,l}$ by
\be
\Fi_{s,l} \equiv \{ (a,b),\ 1\leq a <b \leq s \}
	\cup \{ (c,s+1), \ 1\leq c\leq l  \}
,\ee
such that the following simple relations hold:
\ba
\{ (l+1,s+1),\, \Fi_{s,l} \} &=& \Fi_{s,l+1}
,\\
\Fi_{s,s} &=& \Fi_{s+1,0}
.\ea
Applying formula (\ref{singola}), with $l+1\leq s$, we have
\be
P_{\Fi_{s,l+1}} = P_{ (l+1,s+1), \Fi_{s,l} } =
\widetilde{\cF}_{s+1,1,l+1} (P_{\Fi_{s,l}})
.\ee
By iteration we have the thesis
\be
P_{\Fi_{s+1,0}} = \widetilde{\cF}_{s+1,1,s}\, \ldots \widetilde{\cF}_{s+1,1,2}
\,(P_{\Fi_{s,1}})
\label{optot}                                      
.
\ee
Moreover, by definition of conditional probability we have 
\be
P_{\Fi_{s,1}} = \tilde{P}_{(1,s+1)} \, P_{\Fi_{s,0}}
\label{opgg}
,\ee
where $\tilde{P}_{(1,s+1)}$ is given by~eq.(\ref{prima}).
Therefore we have proven the following

\begin{theorem} \label{th:spillo}
If the 3-replicas overlap distribution $P_{12,23,13}$ is
ultrametric ({\it i.e.}, of the form (\ref{treultra})), and in the limits of 
validity of theorem \ref{thegg},  the overlap distribution  is 
uniquely determined in terms of $P_{12}$.
The explicit form of the distributions of overlaps among $s$ replicas, 
for any $s$ ({\it i.e.}, the $s$-replicas marginals $P_{\Fi_{s,0}}$),
 can be calculated by repeated applications of eqs.~(\ref{optot},\ref{opgg}).
\end{theorem}

Since Theorem~\ref{th:dist}.{$a$} proves the hypothesis of 
Theorem~\ref{th:spillo} in 
the case of  mean field spin glass models, this completes the proof 
of its part {\it b}.

The explicit construction  (\ref{optot}, \ref{opgg}) clearly shows that
ultrametricity and the Ghirlanda--Guerra relations can be considered as 
{\it complementary} in order to determine the full overlap distribution, 
in the sense that one can hold independently of the other, but together 
they determine explicitly the overlap measure in terms of the one-overlap 
distribution.

Results of this kind were obtained by Parisi in the case 
$s=3$~\cite{parisi}.

\section{Conclusions}

It has been shown how mean field disordered models can be successfully
analyzed in a mathematically rigorous framework, with a simple {\it Ansatz}
which is completely different from the Replica Simmetry Breaking {\it Ansatz} .
 In the S.--K. spin glasses case,
the main features of the accepted physical solution 
-- the Parisi solution -- have been obtained. The method exploited, 
due to F.~Guerra, is 
based on the cavity method and general theorems, and can therefore 
be applied to other disordered mean field models such as the multi-spin 
interaction spin glasses or neural networks.

The functional order parameter $x(q)$ has been introduced 
in the S.--K. model. By the {\it Ansatz} that $x$ is indeed the correct order 
parameter, all physical observables have been expressed in terms of it. The 
physical interpretation of the functional order parameter ({\it i.e.} 
$d\, x(q)/dq\; =\; P(q)$) results, and ultrametricity of overlaps is 
derived as a natural consequence of a branching diffusion process. 

It has been shown by explicit construction that 
ultrametricity of the 3-replicas overlap distribution 
together with the Ghirlanda--Guerra relations determines the distribution 
of overlaps among $s$ replicas, for any $s$, in terms of $P_{12}$.

\section*{Acknowledgments}

The authors wish to express their warmest thanks to F.~Guerra,
for fruitful suggestions and stimulating discussions.

\clearpage
\section{Appendix}

We report the explicit expressions of $\rho_s^{{}^{(+)}}
(\{{\mathrm y}_{ab}\})$ for $s = 2, 3,4$. 
For two replicas we have
\be
  \rho_2^{{}^{(+)}}({\mathrm y}_{12}) = 
  \int_0^1 dq\ \rho(q)\ \d ({\mathrm y}_{12} - q)
\ee
for three replicas 
\begin{eqnarray} 
\rho_3^{{}^{(+)}}(\{{\mathrm y}_{ab}\})
  & = & \unsu{2}\ \int_0^1\, dq\ \rho(q)\, x(q) \prod_{(a,b)\subset G_3}
        \d ({\mathrm y}_{ab} - q)\ + \nn \\[0.75em]
  & + & \unsu{2}\ \sum_\pi{}^{(3)}\ \int_0^1\, dq\, \int_q^1\, dq'\ 
        \rho(q)\, \rho({q'})\, \d ({\mathrm y}_{\pi_1\pi_2} - q') 
        \prod_{(a,b)\subset G_3\setminus(\pi_1,\pi_2)}
        \d ({\mathrm y}_{ab} -q) \nn\\
\end{eqnarray} 
and for four replicas
\begin{eqnarray} 
  &   & \rho_4^{{}^{(+)}}(\{{\mathrm y}_{ab}\}) = \nn \\[1.5em]
  & = & \unsu{3}\ \int_0^1\ dq\  
        \rho(q) x^2(q) \prod_{(a,b)\subset G_4}
        \d ({\mathrm y}_{ab} - q)\ + \nn \\[0.75em]
  & + & \unsu{6}\ \sum_\pi{}^{(6)}\ 
        \int_0^1\ dq\ \int_q^1\ dq'\ \rho(q) x(q) \rho({q'})\   
        \d ({\mathrm y}_{\pi_1\pi_2} - q') 
        \prod_{(a,b)\subset G_4\setminus(\pi_1,\pi_2)}
        \d ({\mathrm y}_{ab} -q )\ + \nn\\[0.75em]
  & + & \unsu{6}\ \sum_\pi{}^{(4)}\ 
        \int_0^1\ dq\ \int_q^1\ dq'\ 
        \rho({q}) \rho({q'}) x(q')\ 
        \prod_{(a,b)\subset G_3(\pi_1,\pi_2,\pi_3)} 
        \d ({\mathrm y}_{ab} - q')\ \times \nn \\
&\times&\prod_{(a,b)\subset G_4\setminus G_3(\pi_1,\pi_2,\pi_3)} 
        \d ({\mathrm y}_{ab} - q)\ + \nn \\[0.75em]
  & + & \unsu{6}\ \sum_\pi{}^{(3)}\ 
        \int_0^1\ dq\ \int_q^1\ dq'\ \int_q^1\ dq''\ 
        \rho({q})\rho({q'})\rho(q'')\ 
        \ \d ({\mathrm y}_{\pi_1\pi_2} - q') 
        \d ({\mathrm y}_{\pi_3\pi_4} - q'')\ \times \nn \\
&\times&\prod_{(a,b)\subset G_4\setminus\{(\pi_1,\pi_2),(\pi_3,\pi_4)\}}
        \d ({\mathrm y}_{ab} -q )\ + \nn\\[0.75em]
  & + & \unsu{6}\ \sum_\pi{}^{(12)}\ 
        \int_0^1\ dq\ \int_q^1\ dq'\ \int_{q'}^1\ dq''\ 
        \rho({q})\rho({q'})\rho(q'')\  
        \ \d ({\mathrm y}_{\pi_1\pi_2} - q'')\ \times \nn \\
&\times&\prod_{(a,b)\subset G_3(\pi_1,\pi_2,\pi_3)\setminus(\pi_1,\pi_2)}
        \d ({\mathrm y}_{ab} -q' ) 
        \prod_{(a,b)\subset G_4\setminus G_3(\pi_1,\pi_2,\pi_3)}
        \d ({\mathrm y}_{ab} -q ) \nn\\
\end{eqnarray} 
Here $G_r(i_1,\ldots,i_{r})$ is the complete graph with vertices
$(i_1,\ldots,i_{r})\subseteq \{1,\cdots,s\}$
\footnote{clearly $G_s \equiv G_s (1,\dots,s) = \f_{s,0}$};
$\sum_\pi{}^{(n)}$ indicates the sum on all different $n$ 
permutations $\pi$ on $G_r$ vertices' indexes, which 
render permutation invariant the associated measure. 
The numbers $p_i$, the probabilities of different ultrametric events, 
are obtained by normalizing the corresponding measures;
counting together the permutations of variables they are, 
for three replicas, $(1/4,3/4)$ and for four replicas 
$(1/9,1/6,2/9,1/6,1/3)$. 

The recipe to construct $\rho_s^{{}^{(+)}}(\{{\mathrm y}_{ab}\})$ 
is based on the costruction of abstact trees with a root and 
$s$ ``leaves'', which carry the indices ${\mathrm y}_{ab}$
The $\rho_s^{{}^{(+)}}$  is given by a sum on 
all such trees constructed by elementary branchings: 
each element in the sum is an integral on at most $s-1$ variables 
of the weight $w_T(\ .\ )$ associated with the tree $T$. 
For $T$ given, $w_T$ is the product of the combinatorial factor 
$[(s-1)!]^{-1}$ times the weights of the branchings forming the tree 
\footnote{a branching formed by an input and, say, $i$ outputs, has a weight 
$w_i (q) = (i-2)!\ \rho_q\ x_q{}^{i-2}$} 
and suitable $\theta$ and $\d$ functions on the integral variables 
and the the output variables ${\mathrm y}_{ab}$, 
according to the tree structure. 

A simple way to deduce the number of structurally equivalent graphs, 
goes as follow: we use a scale transformation in (\ref{ide}) to obtain 
the generic term $\Ldi{z}{(s)}{\f,\dots,\f}$
in terms of $\{\Ldi{.}{(s')}{\f,\dots,\f}\}$, for $ 1 \leq s' < s$ in the 
expansion of 
$\Ldi{z}{}{\f} \equiv\ \sum_{}^{}\Ldi{z}{(s)}{\f,\dots,\f}/{s!}$ .  
Let $\f\to\l\ \f$ be this scale transformation:  
it is $\Ldi{z}{{}^{1}}{\f}=\f_z$ and, for $s\geq 2$ 
\be 
 \sum_{s=2}^\infty \frac{\l^s}{s!} \Lf_z^{(s)}\le[\f,\dots,\f\ri] = 
 \sum_{i=2}^\infty \frac{\l^i}{i!} 
            \Op{i}{z}\le[
             \sum_{s_{{}_1}=2}^\infty
              \frac{\l^{s_{{}_1}-1}}{s_{{}_1}!}\Lf_{.}^{(s_{{}_1})}\le[\f\ri]
             ,\dots,
             \sum_{s_i=2}^\infty
              \frac{\l^{s_i-1}}{s_i!}\Lf_{.}^{(s_i)}\le[\f\ri]
             \ri]
 \label{eila} 
\ee 
By multilinearity of the operators, equating terms with equal powers of $\l$,
we have
\be 
 \Lf_z^{(s)}\le[\f,\dots,\f\ri] = 
        \sum_{i=2}^{s} \sum_{\{m_j\}}\!{}' 
        \le(\Op{i}{z}\le[(\Lf^{{}^{(1)}}\le[\f\ri])^{m_{{}_1}},\dots, 
        (\Lf^{(s-1)}\le[\f\ri])^{m_{s-1}}\ri]\ri)_{(S)} 
 \label{allsim} 
\ee 
the sum $\sum'_{\{m_j\}}$ is on all $m_j \geq 0$ 
with the bounds $\sum_j m_j = i$ and $\sum_j jm_j = s$; 
$(\Lf_j [\f])^{m_j}$ is briefly for 
$m_j$ repetitions of $\Lf_j [\f]$  operator as argument of 
$\Op{i}{z}$ and finally ${(S)}$ is the symmetrical factor 
in the $\f$'s given by 
\be 
  S = \frac{s!}{m_1!\cdots m_{s-1}!\ 2!^{m_2}\cdots (s-1)!^{m_{s-1}}} 
\ee 
which counts all structurally equivalent graphs. 


\clearpage
\newcommand {\vol} [1] {{\bf #1}}



\begin {thebibliography}{99}

\bibitem  {SK} S. Kirkpatrick, D. Sherrington,
          {\it Infinite-ranged models of spin-glasses}
          Phys. Rev. B \vol{17}, 4384, (1978)
\bibitem  {PMV} M. Mezard, G. Parisi, M.A. Virasoro,
          {\it Spin Glass Theory and Beyond}
          World Scientific, (1986)
\bibitem  {pash} L. Pastur, M. Shcherbina, {\it The absence of self-averaging
		of the order parameter in the Sherrington--Kirkpatrick model},
	  J. Stat. Phys., \vol{62},1, (1992)
\bibitem  {shch} M. Shcherbina, {\it On the replica symmetric solution for the 
	  Sherrington-Kirkpatrick model}, Helv. Phys. Acta, \vol{70},838-853, 
	  (1997)
\bibitem  {gue1} F. Guerra,
          {\it Fluctuation and Thermodynamic variables in
               Mean Field Glass } in ``Stochastic Processes, Physics
          and Geometry, II'', S. Albeverio {\it et al.}, eds., Singapore
          (1995); also avalaible through http://romagtc.roma1.infn.it
\bibitem  {gue2} F. Guerra,
          {\it Functional Order Parameters for the Quenched Free
               in Mean Field Spin Glass Models} in ``Field Theory
          and Collective Phenomena'', S. De Lillo {\it et al.}, eds.,
          Singapore (1995)
\bibitem  {gue3} F. Guerra,
          {\it The Cavity Method in
               the Mean Field Spin Glass Model} in
               ``Advances in Dynamical Systems and Quantum Physics''
          S. Albeverio {\it et al.}, eds., Singapore (1995)
\bibitem  {gue4} F. Guerra,
          {\it About the Overlap Distribution in
               Mean Field Spin Glass Models} Int. Jou. Mod. Phys.
          B \vol{10}, 1675--1684 (1996)
\bibitem  {tala} M. Talagrand, {\it The Sherrington Kirkpatrick model~:
          a challenge for mathematicians}, Prob. and Rel. Fields, 
          \vol {110}, 109-176 (1998)
          and M. Talagrand, {\it Replica symmetry breaking and exponential
          inequalities for the Sherrington Kirkpatrick model}, Ann. Prob., 
          to appear
\bibitem  {gueun} F. Guerra, {\it On the mean field spin glass model},
	  in preparation
\bibitem  {ghigu} S. Ghirlanda, F. Guerra, {\it General properties of 
          overlap probability distributions in disordered spin systems. 
          Toward Parisi ultrametricity.}, J. Phys. A, \vol{31}, 
          9149-9155 (1998), and 
	  S. Ghirlanda, {\it Sulle propriet\`a ultrametriche degli stati di 
	  sistemi disordinati}, Thesis at Universit\`a di Roma
	  ``La Sapienza'' (in Italian), (December 1996)
\bibitem  {parisi} G. Parisi, {\it On the probabilistic formulation of the 
	  replica approach to spin glasses}, cond-mat/9801081, (1998)
\bibitem  {AC} M. Aizenmann, P. Contucci, {\it On the stability of
	  the quenched state in mean field spin glass models.},
 	  J. Stat. Phys. \vol{92}, 765-783 (1998) 
\bibitem  {num1}
	  E. Marinari, G. Parisi and J. J. Ruiz-Lorenzo,
	  {\em Numerical Simulations of Spin Glass Systems},
	  in ``Spin Glasses and Random Fields''
	  A.~P.~Young ed., World Scientific, Singapore (1998), and
 	  cond-mat/9701016;
	  A.~Billoire and E.~Marinari, cond-mat/9910352.
\bibitem  {Parisi_rev}
	  E.~Marinari, G.~Parisi, F.~Ricci Tersenghi, J.~Ruiz Lorenzo 
	  and F.~Zuliani,
	  {\em Replica Symmetry Breaking in Short Range Spin Glasses: A Review
	  of the Theoretical Foundations and of the Numerical Evidence},
	  to be published on J. Stat. Physics, cond-mat/9906076.             

\end {thebibliography}
\end{document}